# Analysing Human Interaction with Electronic Displays in Microgravity

Pradipta Biswas, Himanshu Vishwakarma, Mukund Mitra, KamalPreet Singh Saluja, Aumkar Kishore Shah, Indian Institute of Science, Bangalore, India

Corresponding Author's email: pradipta@iisc.ac.in

***Abstract:*** Human Space Flight missions often require interaction with touchscreen displays. This paper presents a study of investigating human machine interaction with touchscreen using both finger and stylus in the International Space Station. The study also reports cognitive state of astronauts in the form of spatial 2-back test and mental well-being through self-reported scales. We presented a series of results comparing pointing and selection performance among ISS crews, ground crews and university students, finger-based touching and stylus-based touching in microgravity and mental well-being scores. We reported that finger-based pointing is statistically significantly faster than stylus-based pointing in microgravity based on analysis of 420 pointing tasks in ISS from 2 astronauts. We also did not find any significant difference among pointing performance and mental state of astronauts and students on ground. Results from the study can be used to predict pointing and selection time from dimension and position of GUI (Graphical User Interface) elements for cockpits of spacecraft.

## Introduction

Rapid aiming movements were analysed from last decade of 18th century. Presently, ISO 9241 pointing task [ISO 9241-303, 2011] inspired by a paper published in 1954 by Paul Fitts is used for analysing pointing devices including touchscreen devices. In the context of future deep space human space flight programs, astronauts were supposed to operate spacecraft, control docking arm and teleoperate rover during space flight missions – a large portion of these interactions will be performed using touchscreen devices in microgravity. On earth, microgravity involving human participants is simulated through parabolic flight. While glass cockpits and touchscreen-enabled Multi-Function Displays (MFDs) are already integrated in crewed aviation platforms, space shuttle cockpits were slow in integrating MFDs and touchscreen systems. Recent SpaceX crew modules are operated by touchscreen or virtual buttons. The proposed study investigated

 a. Effect of microgravity on interaction with electronic screens
 b. Analysing pointing tasks under microgravity-how the rapid aiming movement adjusts itself under microgravity
 c. Effect of ISS environment on mental well being

through a series of web based bespoke applications. The tests involved a pointing trial requiring participants to repeatedly tap on a white box on an iPad using fingers and Apple Pencil, an online cognitive test involving remembering position of a red square in 3×3 grid and filling up two online questionnaires on general (not clinical) mental well-being. The present study in ISS will be used to develop design guidelines for developing cockpits of future spacecrafts and human habitats in microgravity environments like space station, lunar or planetary surfaces.

## Literature Review

We act in various ways of moving our various body parts like walking, running, nodding head, hand gesturing and so on. In the field of user interaction, we concentrate on a particular type of movement known as rapid aiming movement. Analysis of rapid aiming movement has a rich history starting from last decade of 18th century, when Woodworth [Eliott 2001] was inspired by the speed and accuracy of hammering activity of construction workers and concluded that with visual feedback people make more error with higher speed of execution while without visual feedback, error does not depend on accuracy. The finding leads to two phases of rapid aiming movement – the ballistic phase that does not depend on visual feedback and the homing phase that depends on visual feedback. Later in 1954, Paul M Fitts' published a seminal paper where he predicted the task completion time as a function distance to target and width of target [Fitts 1954]. In 1979, Stuart Card used Fitts' Law to predict mouse as the most efficient pointing device compared to joystick and head movement tracker based on which Xerox Parc and later other companies started to market computer mouse. Later the Fitts' Law task became part of ISO 9241 standard [ISO 9241-303, 2011] and presently used to compare or evaluate any pointing device. Besides numerous terrestrial applications, Fitts' Law study was also conducted at the International Space Station [Holden 2023]. Holden's study was an elaborate well-designed study involving seven astronauts and four tasks spanning more than three months. For pointing tasks, the study used two different target sizes. The study did not find any significant difference in pointing performance between ISS and ground crews after first week of stay in the ISS. Ciofani [2012] reported results from a Fitts' Law task undertaken during parabolic flight at both hypergravity (1.8G) and microgravity from four participants. Authors reported more variance in movement times in microgravity and hypergravity conditions compared to control condition of 1G. Authors fit a third order polynomial between movement time (MT) and Index of Difficulty (ID) for hypergravity and microgravity conditions. Bieg [2024] reported results from a touchscreen tapping task at a head-down tilt (HDT) bed rest as part of the Artificial Gravity Bed Rest, European Space Agency (AGBRESA) at the German Aerospace Center (DLR) in Cologne. Researchers reported validity of Fitts' Law in HDT condition in terms of regression parameters between movement time and Index of Difficulty (ID) with slopes ranging from 143 to 163 and $R^2$ ranging from 0.94 to 0.96 using data collected from 24 healthy participants aged 24 to 55 years.

## Method

Following the literature survey and previous study undertaken at the International Space Station, we aimed to contribute to developing and evaluating graphical user interfaces for space missions. We reported raw data in terms of task completion times with respect to target size and distances from centre of screen and N-back test that can be used to evaluate task completion times using GOMS model. Three ground crews, who were test pilots and astronaut designate, and two university students also undertook the tests during the study period. We included both ground crews and students to help in setting sampling criteria during early-stage evaluation of spacecraft HMI as trained test pilots are often rare and many space programs already started training civilians as astronauts. Our study consisted of the following five tests to be conducted once per day during 14 days stay period at the International Space Station.

1. Pointing task with Finger
2. Pointing Task with Stylus
3. Spatial 2-Back Test [Gazzaniga 2018]
4. Perceived Stress Test [Cohen 1983; 1988]
5. Mental Well-being Test [WHO 2025]

The first two tests aimed to evaluate fine motor skill related to human computer interaction. The third test was an objective evaluation of cognitive status while the last two tests were online self-reported evaluation. The study took NASA IRB approval and none of the tests was intended to collect or evaluate any physiological parameters or other Personal Identifiable Information according to ISO 27018:2019 standard. All tasks were undertaken on an Apple iPad Pad Pro (11-inch) (3rd generation) with default screen settings in Landscape mode and an Apple Pencil (version 2.0) was used as stylus for the second pointing task. In the following paragraphs, these tests are described in further details:

**Pointing Tasks:** The tasks were designed following ISO 9241 pointing task using three different target sizes and distances

- ➢ Sizes: 20 pixels, 35 pixels and 50 pixels (5, 10, 15mm)
- ➢ Distances from centre of screen: 120 pixels, 140 pixels and 160 pixels (24, 27.5, 32mm)

The order of the pointing tasks with finger and stylus were counterbalanced in subsequent days to minimize order effect.

**Cognitive Task:** We undertook a 2-back Spatial test where one of nine grids were highlighted for 1 seconds and participants were instructed to press a match button when the highlighted grid position matched the previous to previous (2-back) highlighted grid position. During each session, there were 22 iterations of grid highlighting, the order of highlights was random. We measured the total numbers of correct, wrong and missed selections.

**Perceived Stress and Mental Well-being Tests:** We developed an online form to collect data on standard questionnaire. The perceived stress scale had 10 questions while the WHO test had 5 questions. Each question was answered in a 5-points Likert scale. Score calculation was undertaken according to respective manuals [Cohen 1983, WHO 2025].

## Results

One astronaut undertook the task 11 days while a second astronaut could undertake the trial only for two days. Table 1 below shows the number of sessions recorded from each participant. It may be noted that one session of the pointing task consisted of 60 pointing tasks, and one round of execution of cognitive, perceived stress and mental well-being tests. For example, row 2 indicates, Astronaut 2 undertook 60 pointing tasks with finger, 180 pointing tasks with stylus, and undertook the 2-back Spatial task twice, filled up the Perceived Stress questionnaire once and WHO 5 questionnaire twice. Due to the uneven nature of the data in the form of sessions we need to use a subset of whole data for subsequent analysis. We undertook the following analysis among ISS crews, ground crews and students in this paper

- ➢ Comparison of pointing times by finger and stylus
- ➢ Comparison of results of 2-Back tests among ISS crews, ground crews and students
- ➢ Comparison of results of self-reported perceived stress and mental wellbeing among ISS crews, ground crews and students

High definition version of all graphs can be found at supplement AX-4 CSIR Presentation.pptx.

**Table 1** Recorded session from Participants

| Participants | Finger Pointing | Stylus Pointing | 2-BackSpatial | Perceived Stress | Mental Wellbeing |
|---|---|---|---|---|---|
| *Astronaut 1* | 6 | 9 | 13 | 10 | 9 |
| *Astronaut 2* | 1 | 3 | 2 | 1 | 2 |
| *Ground Crew 1* | 8 | 7 | 8 | 8 | 7 |
| *Ground Crew 2* | 3 | 3 | 3 | 3 | 3 |
| *Ground Crew 3* | 3 | 3 | 3 | 3 | 3 |
| *Student 1* | 8 | 8 | 8 | 8 | 8 |
| *Student 2* | 7 | 7 | 7 | 7 | 7 |

For comparing pointing times, first we calculated descriptive statistics and presented a box-plot of the pointing times in different conditions (Figure 1). Next, we removed pointing times exceeding two seconds considering those outliers. The value of two seconds was above outer fence values as well for all conditions. We averaged the pointing and selection times with respect to each combination of target size and distances. The total number of pointing tasks were more than 400 for each condition. Figure 2 shows the Pointing Time vs Index of Difficulty (ID) plot. It may be noted here that the ID is calculated from the 2D screen, not the actual ID considering the distance of user from screen.

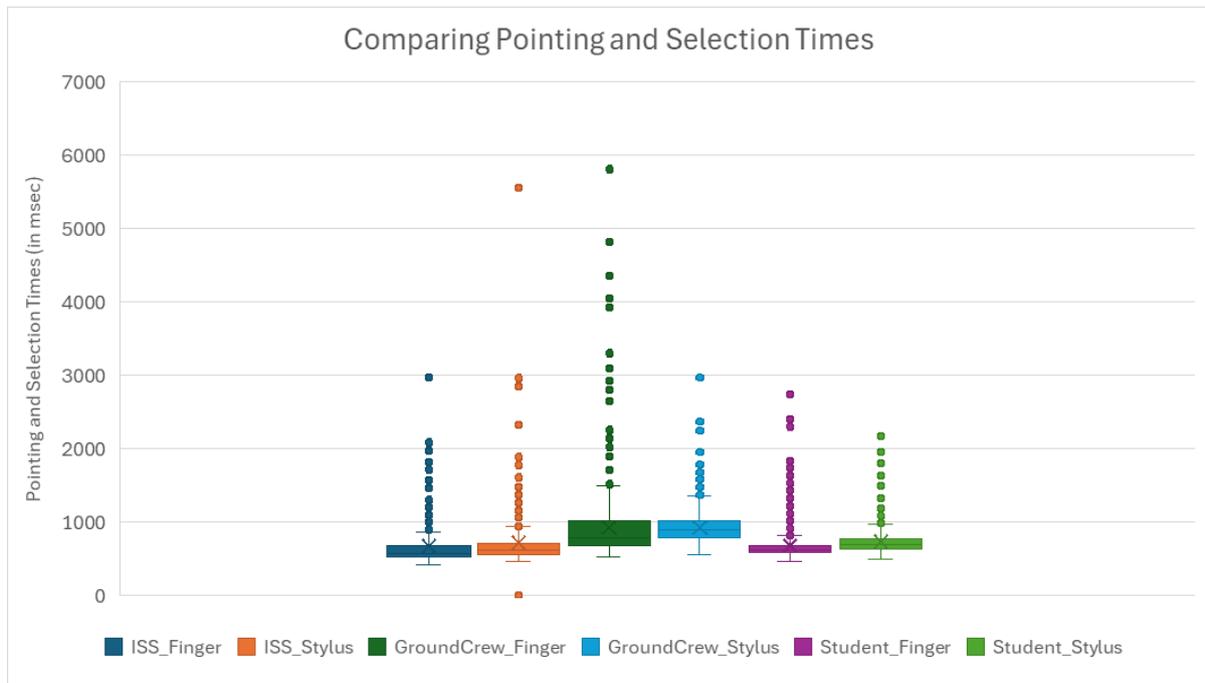

**Figure 1.** Box Plot of Pointing and Selection Times for different conditions

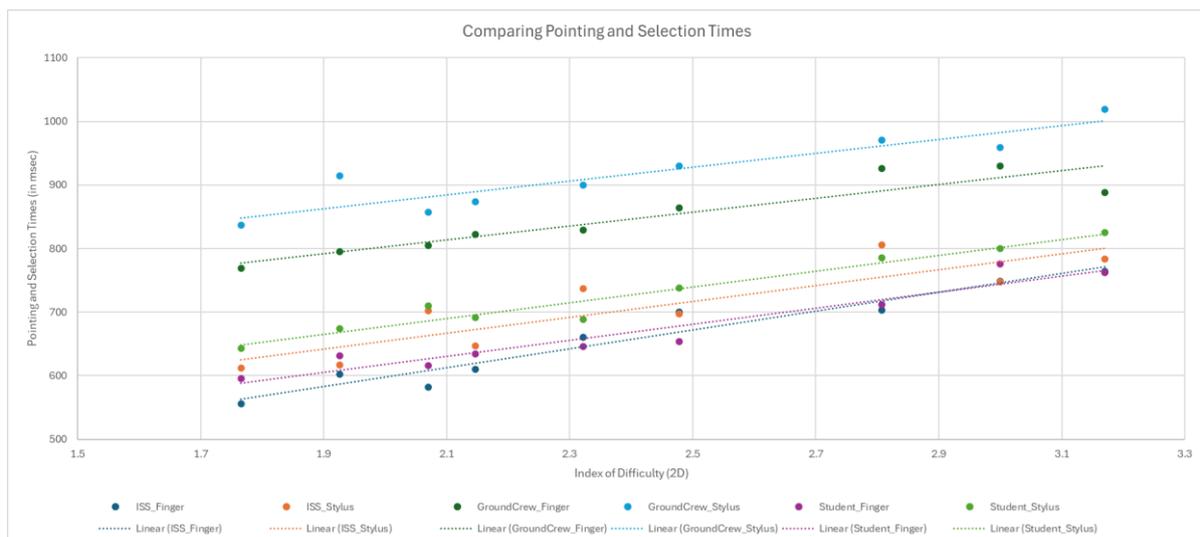

**Figure 2.** ID vs MT plot for different conditions

We found a statistically significant difference between pointing by finger and stylus $F(1,24)=76.32$, $p<0.001$, $\eta^2=0.76$. We also found statistically significant difference between ground crews and two other groups of users (ISS crews and students). Table 2 below furnishes the correlation between ID (2D) and pointing and selection times.

**Table 2.** Correlation between ID and Pointing and Selection Times

| ISS_Finger | ISS_Stylus | GroundCrew_Finger | GroundCrew_Stylus | Student_Finger | Student_Stylus |
|---:|---:|---:|---:|---:|---:|
| 0.97 | 0.88 | 0.93 | 0.92 | 0.96 | 0.98 |

The following equations can be used to calculate pointing and selection times for different sizes and positions of targets on a Graphical User Interface **in microgravity environment**

**Pointing time with Finger (in msec) =** $148.15 \, Log2(1 + \frac{Distance\ to\ Target}{Width\ of\ Target})$ **+ 301.61**

**Pointing time with Stylus (in msec) =** $124.58 \, Log2(1 + \frac{Distance\ to\ Target}{Width\ of\ Target})$ **+ 405.33**

We did not find any statistically significant difference in pointing and selection times between astronauts at ISS and students on ground. The data collection software was also programmed to log cursor movements on screen. As the participants pointed using finger or stylus, which approached from the z-axis, the system did not log the entire trajectory but only a small segment when the finger or stylus was near the screen and mostly over the target region. Based on this limited movement trajectory during the homing phase, we calculated accuracy measures defined by [MacKenzie 2001]. We analysed these parameters between ISS crew and students and found larger values for all accuracy measures except Target ReEntry (TRE), which was recorded as zero for all users and conditions. For ISS crew compared to students, TAC (Target Axis Crossing), ODC (Orthogonal Direction Change), MV (Movement Variability), ME (Movement Error), MO (Movement Offset) were statistically significantly higher at p<0.01 in an unequal variance t-test (figure3). It may be argued that the difference may be due to different finger movement patterns at the final stage of tapping or unknown difference in settings of iPads used by ISS crew and students. We separately compared the accuracy measures only for pointing tasks involving the highest values of IDs and found that the difference is bigger than taking all pointing tasks together (figure4). We also got similar results with stylus-based pointing. The results indicate that even though the overall pointing and selection times were not different between ISS crews and students but there may be subtle differences in fine adjustments of finger or stylus movement at the homing phase due to microgravity, which may be further analysed in future with accurate finger movement trackers.

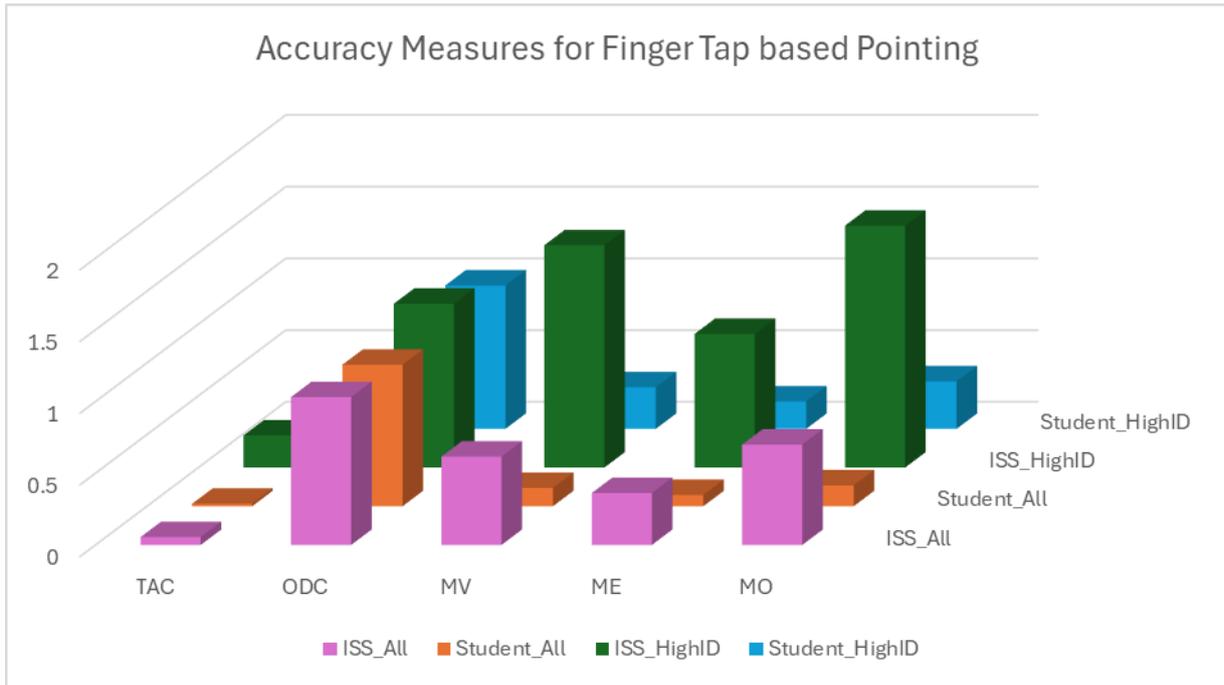

**Figure 3.** Comparing Accuracy Measures between ISS Crew and Students and between All Pointing Tasks and Tasks involving Highest Value of ID

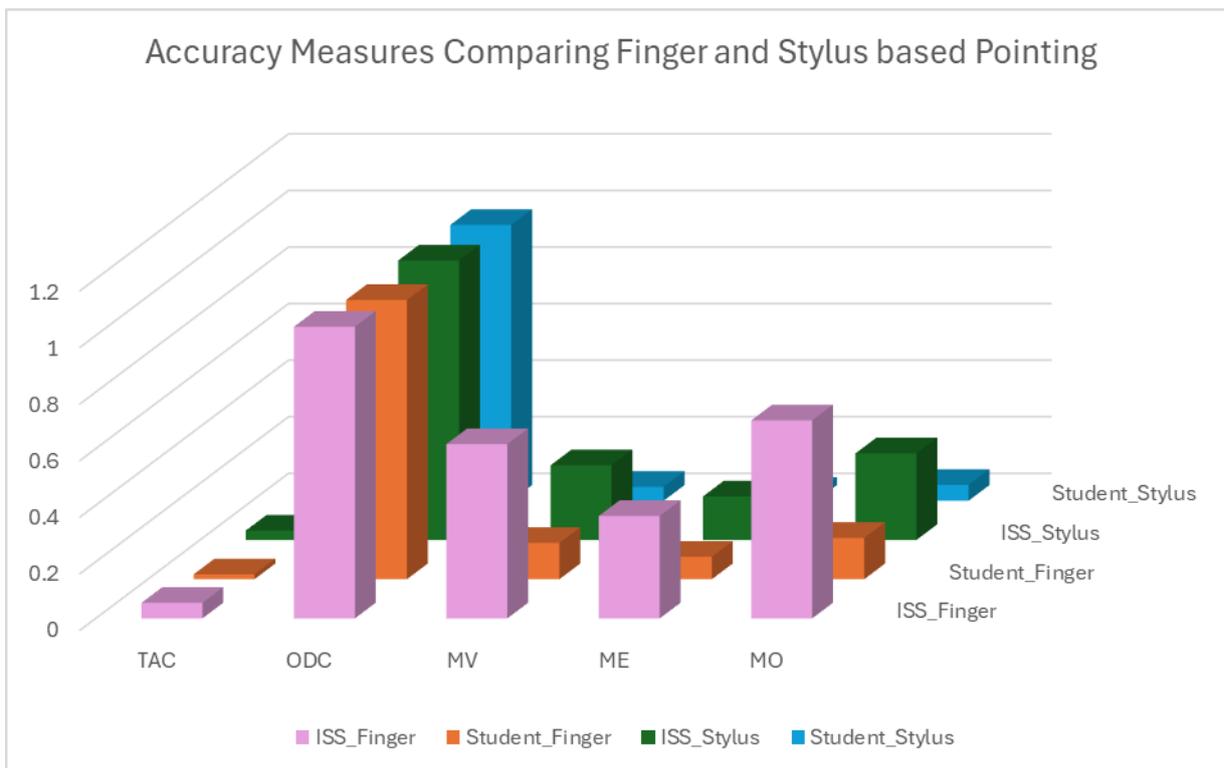

**Figure 4.** Comparing Accuracy Measures between ISS Crew and Students for Finger and Stylus based Pointing

In terms of 2-back spatial task, we plotted the number of correct, wrong and missed selections by astronaut 1 (who undertook it 13 times over his stay duration at ISS), Ground Crew 1 and two students. We did not use data from Astronaut 2, who only undertook the task twice and ground crews 2 and 3, who undertook the task 3 times. The data is plotted in chronological order (figure 5).

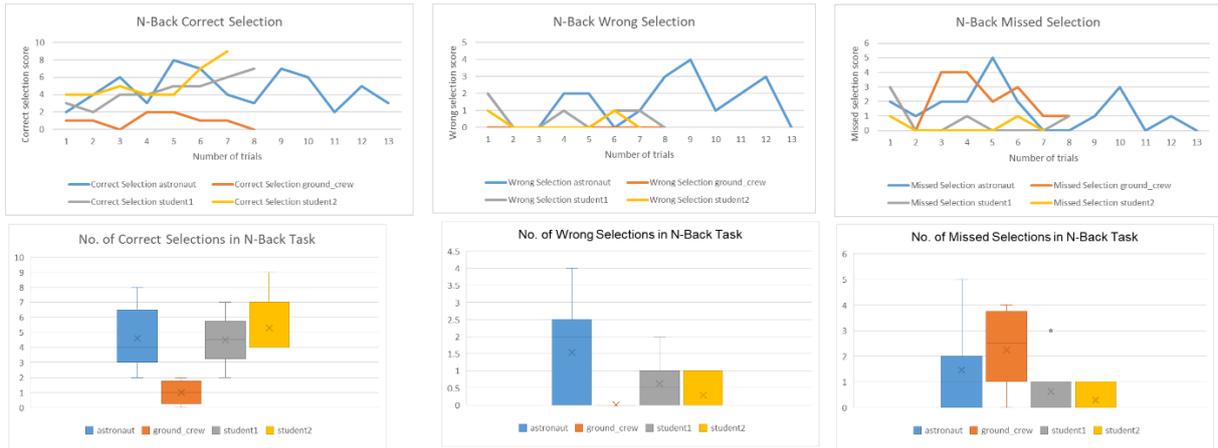

**Figure 5.** Results from N-back Task

- It may be noted from the graph that different strategies were employed by ISS and Ground Crews
    - ISS crew's performance was similar to students, who tried to have more correct selections at a cost of more wrong selections. and we did not find any significant difference among students and astronaut1's performance in Mann Whittney U-test.
    - Ground crew1 tried to minimize wrong selection, had 1 correct selection and 2.25 missed selections on average. Although we did not plot in figure 5, ground crews 2 and 3 had only 3 sessions each and had nearly 4 correct selections on average and did not miss any selection.

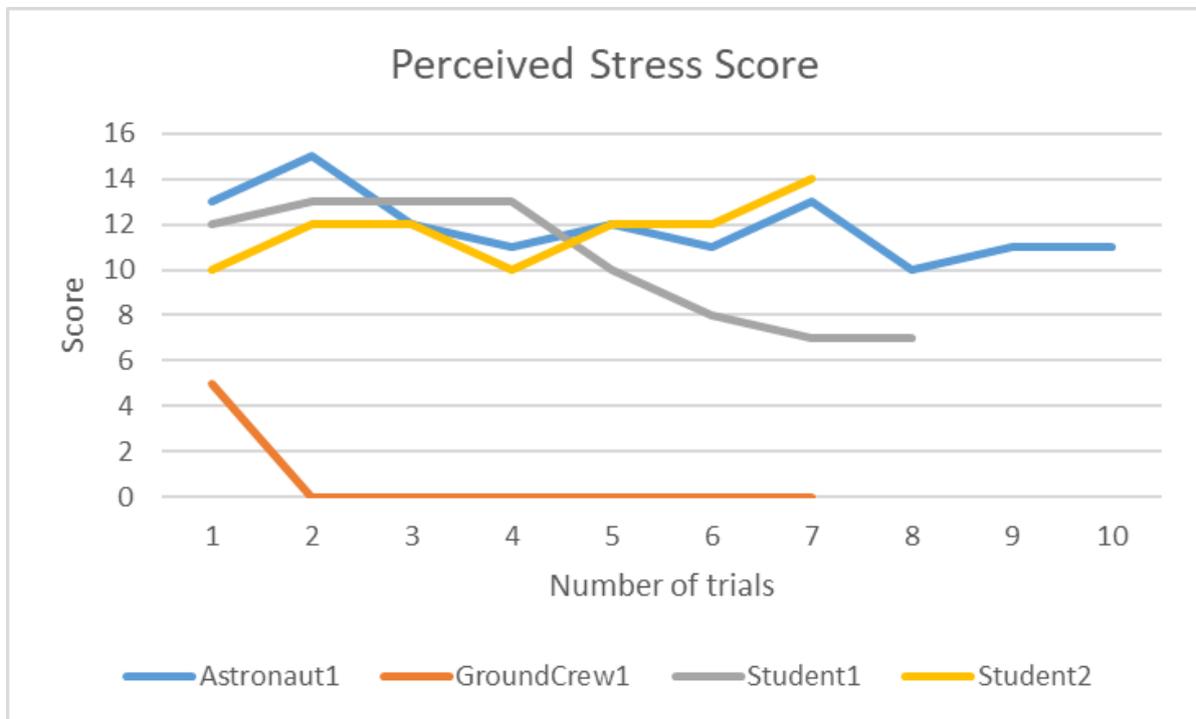

**Figure 6.** Self-reported Perceived Stress Scale

The self-reported scales (figure 6) show a flat response (same score) from the ground crew 1 while the ISS crew reported higher scores than students indicating no sign of mental stress or discomfort. The graph (figure 6) illustrates changes in perceived stress scores over repeated trials. Three participants started with stress levels (~10–14), which falls within the low-moderate stress range based on Cohen's PSS interpretation. Over trials, Astronaut 1 showed a gradual decline in stress levels, indicating possible adaptation or habituation to the task, while the other two showed fluctuating scores, reflecting variable stress responses to changing conditions or task demands. One participant starts at a low score (~5) and rapidly drops to zero, suggesting minimal perceived stress, or possibly disengagement or misunderstanding of current task. The repeated exposure to the task resulted in decreased or stabilized perceived stress for most participants, which is consistent with the adaptation theory according to which there is habituation to repeated stimuli and decreased responses over time [Thompson 1966]. Cohen [1988] reported that moderate levels are common but can shift with context, coping mechanisms, and repeated exposure are also reflected. The perceived stress tends to decrease or fluctuate within a low-moderate range over repeated trials, which may imply participants adapting to the task or employing coping strategies as they become more familiar with it.

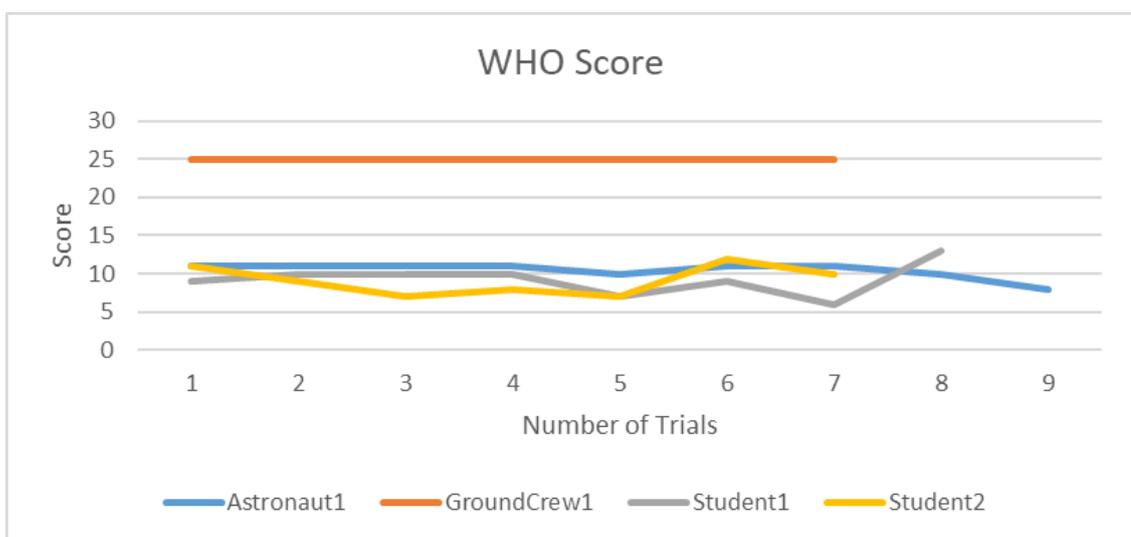

**Figure 7.** Self-reported Mental Well-being Scale

For the WHO scale (figure 7), Astronaut1 displayed moderate WHO-5, along with high perceived stress early in the trials that tapered later. Behaviourally, they had the highest correct responses but also the highest wrong selections, which indicates an impulsive, speed-over-accuracy strategy [Jaeggi 2010]. The decline in WHO-5 across trials may reflect cumulative cognitive strain [van der Linden 2003], possibly driven by their high engagement and error-prone pace. The Ground Crew reported the highest possible WHO-5 scores (~25) throughout the trials, indicating high subjective well-being [Topp et al., 2015] and also reported near-zero stress from the second trial onward. Student 1 showed a low and somewhat declining WHO-5 score, high initial stress that decreased over time, and accurate task performance with low false alarms. This profile suggests an effective adaptation to the task despite low baseline well-being, consistent with findings that attentional control can be maintained under moderate stress in some individuals [Matthews 2002]. Student 2 had a moderate and variable WHO-5 score and generally stable stress levels Performance was balanced, with low false alarms and few misses, indicating good cognitive control. The stress spike suggests reactivity to specific task demands or fatigue [Hockey 1997].

In terms of subjective preference expressed during post-session feedback, ISS crew reported ease of use of all tasks and preferred pencil over tap by finger. Ground crews reported a few occasions of multiple taps for selecting small targets on screen using fingers.

## Discussion

The main contributions from this study are

- An absolute estimation of pointing and selection times with touchscreen with respect to 9 different target size and distances that can be directly used to compute task completion times with touchscreen for various tasks like emergency response, manual settings change and even piloting the dragon crew using touchscreen. ISS crews could select target of sizes between 5 and 15 mm within 500 to 900 msecs.
- Confirmation of validity of Fitts' Law in terms on high correlation between pointing and selection times and Index of Difficulty in microgravity environment.
- A comparison between touchscreen and stylus in microgravity environment found that finger touching was statistically significantly faster than stylus.
- Comparison on cognitive functionality and mental well-being between ISS and Ground crews that can be used to developing training program and task schedule of crews.
- Confirmation of a previous finding that after sufficient ground-based training and one day of acclimatization in microgravity environment, crews' fine motor skill relevant to touchscreen interaction do not degrade compared to ground crews.

The integration of subjective (WHO-5, Perceived Stress) and objective (N-back) measures reveals complex relationships between mental well-being, perceived stress, and cognitive performance. While high self-reported well-being might intuitively predict better cognitive engagement, this was not the case for the ground crew, whose behavioural data suggested disengagement. Conversely, participants with lower well-being scores maintained high accuracy (students), indicating potential resilience. These findings underscore the importance of multi-modal assessment when evaluating cognitive readiness and psychological functioning, particularly in operational or high-stakes contexts such as spaceflight. Reliance on self-report measures alone could lead to erroneous conclusions about an individual's readiness or performance potential. Future work will investigate the situational and motivational factors that drive such divergences and consider integrating momentary (state-level) well-being assessments alongside trait-level measures. The main limitation of the study was the small sample size. The study was designed in context of a private mission to ISS and the IRB approval was limited to only crews involved with the study. Although, two crew could devote time on the study, the main results on pointing and selection times were reported analysing more than 400 pointing tasks and also conform to previous study. The perceived stress and mental well-being study shows

- The strategy of undertaking pointing task (speed vs accuracy trade off) was mainly responsible behind significant difference in pointing times between ISS and ground crews. The involvement of students showed that microgravity did not produce faster pointing performance, but ground crew undertook pointing tasks at a slower rate, most probably to keep it accurate as also indicated by the number of wrong selections in the N-back test results.

- The subjective feedback may not necessarily conform to objective performance, although crew preferred stylus, but they undertook the pointing task faster with finger tapping.

- **It is important to note the study did not intend or should be used as a metric of performance or medical condition evaluation of participating members.**

# Conclusions

This paper presents a study of investigating human machine interaction with touchscreen display in microgravity environment along with reporting cognitive state and mental well-being of astronauts. The study involved pointing tasks with 9 different combinations of target size and distances using finger and stylus, 2-back spatial test, perceived stress and WHO mental well-being tests. Two astronauts, three ground crews and 2 university students took part in the study during 14 days stay at the International Space Station. Our study did not find any significant difference in fine motor skill and associated pointing performance and mental states of ISS crews and students on ground. We found significant difference in pointing performance between finger and stylus in microgravity. We also found that ground crews took significantly longer duration to point and select compared to ISS crews showing difference in strategy or briefing makes a bigger difference than microgravity. Results from the study can be used to design cockpit or human robot interfaces for future space applications [Supplement VoyagerDispaly.mp4].

# Declarations

*Financial Declaration*

The study did not receive funding from any source

*Human Ethics and Consent to Participate declarations*

The study was approved by NASA IRB (STUDY00000851: ANALYSING HMI IN MICROGRAVITY Initial Approval: 1/17/2025). Informed Consents were obtained from all participants.

*Competing Interest*

Authors do not declare any competing interest.

*Author Contribution*

- PB supervised the study and analysed results
- HV, MM and KPS implemented the web-based application for data collection
- AS analysed results of Cognitive and mental well-being tests

# Acknowledgement

This study was conducted as part of the experiment titled 'Analyzing Human Interaction with Electronic Displays in Microgravity' selected by ISRO's Human Space Flight Center (HSFC) for ISRO-NASA joint mission to the ISS (Axiom-4 mission). Authors are grateful to HSFC/ISRO for enabling this study during the Axiom-4 mission.